\documentclass[preprint,showpacs,showkeys,preprintnumbers]{revtex4}

\usepackage{amsmath,amssymb,amscd,amsbsy,amsgen,amsopn,amstext,amsxtra}
\usepackage[mathscr]{eucal}

\begin{document}

\renewcommand{\thefootnote}{\fnsymbol{footnote}}

\preprint{NUP-A-2005-3}

\setlength{\baselineskip}{6.4mm}

~

%\vspace{5mm}

%\title{\Large A covariant gauge-fixing for Yang--Mills theory \\ on sphere}

\title{\Large A BRST gauge-fixing procedure for Yang--Mills theory on sphere}

\author{~\\{\large Rabin Banerjee}\footnote{Email address: 
rabin@bose.res.in}}

\address{S.N. Bose National Centre for Basic Sciences, 
JD Block, Sector III, Salt Lake City, Kolkata 700 098, India}

\author{{\large Shinichi Deguchi}\footnote{Email address: 
deguchi@phys.cst.nihon-u.ac.jp }}

\address{Institute of Quantum Science, College of Science and Technology, 
Nihon University, Chiyoda-ku, Tokyo 101-8308, Japan\\}

%\vspace{-9.67mm}
%\vspace{-3mm}

\begin{abstract}
\setlength{\baselineskip}{6.5mm}
A gauge-fixing procedure for the Yang--Mills theory on an $n$-dimensional 
sphere (or a hypersphere) is discussed in a systematic manner. 
We claim that Adler's gauge-fixing condition used in massless Euclidean QED on a 
hypersphere is not conventional because of the presence of an extra free index,  
and hence is unfavorable for the gauge-fixing procedure based 
on the BRST invariance principle (or simply BRST gauge-fixing procedure). 
Choosing a suitable gauge condition, which is proved 
to be equivalent to a generalization of Adler's condition, 
we apply the BRST gauge-fixing procedure to the Yang--Mills theory on a hypersphere 
to obtain consistent results. 
Field equations for the Yang--Mills field and associated fields 
are derived in manifestly O($n+1$) covariant or invariant forms. 
In the large radius limit, these equations reproduce the corresponding field equations
defined on the $n$-dimensional flat space.
\end{abstract}



%\draft

\maketitle

\newpage

\section{Introduction}
Manifestly O($n+1$)-covariant formulation of gauge theories on an $n$-dimensional 
sphere (or a hypersphere) has been discussed 
in several contexts \cite{Adl,DS,Sho,JR,Ore,BP,NS,Ban}. 
Such a formulation was first developed by Adler to study massless  
Euclidean QED (quantum electrodynamics)  
on a hypersphere in 5-dimensional Euclidean space \cite{Adl}.
Infrared-finiteness of this theory was pointed out there as one of the 
advantages due to compactification of space-time.   
Further study of massless Euclidean QED on a hypersphere was made 
in a manifestly O($n+1$)-covariant way by using the dimensional regularization \cite{DS} 
and it was extended to the case of Yang--Mills theory \cite{Sho}.  
The manifestly O(5)-covariant formulation was also applied to analyzing  
classical and semi-classical behaviors of the pseudoparticle solution in the SU(2) 
Yang--Mills theory \cite{JR,Ore,BP}. 
The connection between the axial anomaly and the Atiyah-Singer 
index theorem was illustrated with the manifestly O($n+1$)-covariant 
formulation in the cases of $n=2$ and $n=4$ \cite{NS}.  
Recently, the manifestly O($n+1$)-covariant formulation was reconsidered 
with the aid of conformal Killing vectors \cite{Ban}. 
There, in addition to the Yang--Mills theory, 
a rank-2 antisymmetric tensor gauge theory and the case involving 
spinor fields were discussed by using formulas derived with the aid of 
conformal Killing vectors 
and spinors.

To study the quantum-theoretical properties of a gauge theory, 
we have to introduce a suitable gauge-fixing condition to this theory. 
In the study of massless Euclidean QED on a hypersphere, 
Adler adopted a gauge-fixing condition appropriate for manifestly 
O($n+1$)-covariant analysis \cite{Adl}.  
This condition makes the analysis quite simple and is applicable 
to higher-dimensions and to the case of Yang--Mills theory. 
However, Adler's condition is unusual in the sense that
it has an extra free index. For this reason, 
Adler's condition is not favorable for the gauge-fixing procedure based on the 
Becchi--Rouet--Stora--Tyutin (BRST) invariance principle  
(or simply BRST gauge-fixing procedure) proposed by Kugo and Uehara \cite{KU,NO}. 
In fact, Adler's condition has not been treated in connection with  
the first-order formalism of gauge-fixing \cite{NL} and with the BRST symmetry 
\cite{NO,KO}.

The purpose of this paper is to apply the BRST gauge-fixing procedure 
to the Yang--Mills theory on a hypersphere in such a way that 
manifestly O($n+1$)-covariance of the theory is maintained.  
To this end, we propose a gauge-fixing condition adapted for 
the BRST gauge-fixing procedure. 
Our condition is equivalent to a generalization of Adler's condition, 
but has no extra free indices. 
A desirable gauge-fixing term is thus defined at the action level,  
and it is shown, with the aid of conformal Killing vectors, 
that this gauge-fixing term yields results compatible with 
those for the Yang--Mills theory on the flat space.

The paper is organized as follows: 
Section 2 is devoted to a brief review of manifestly O($n+1$)-covariant 
formulation of the Yang--Mills theory on a hypersphere.  
Section 3 discusses the gauge-fixing conditions. Here   
the equivalence of a generalization of Adler's condition and 
our gauge-fixing condition without extra free indices is proved.  
Section 4 treats the BRST gauge-fixing procedure for 
the Yang--Mills theory on a hypersphere. 
In section 5, field equations for 
the Yang--Mills field and associated fields 
are derived in manifestly O($n+1$) covariant or invariant forms. 
They are shown to reproduce, in the large radius limit, 
the corresponding field equations defined on the $n$-dimensional 
flat space. Section 6 contains concluding remarks.

\section{Yang--Mills theory on hypersphere}

In this section, we briefly review a manifestly O($n+1$)-covariant formulation of 
the Yang--Mills theory on a hypersphere. 
Following the literature \cite{Adl,JR,Ore,NS,Ban}, 
we consider an $n$-dimensional unit sphere (or a unit hypersphere) $S_{1}^n$  
embedded in $(n+1)$-dimensional Euclidean space ${\bf R}^{n+1}$. 
The hypersphere $S_{1}^n$ is characterized by the constraint $r_a r_a=1$ imposed on  
Cartesian coordinates $(r_a)$ $(a=1,2,\ldots, n+1)$ on ${\bf R}^{n+1}$. 
One may use $(r_\mu)$  $(\mu=1,2,\ldots, n\,;\,0\leq r_\mu r_\mu \leq 1)$ as local 
coordinates on $S_{1}^n$, 
treating $r_{n+1}=\pm\sqrt{1-r_{\mu} r_{\mu}}$ as a dependent variable. 
In this treatment, the angular momentum operators $L_{ab}$ read 
\begin{align}
L_{\mu\nu}&=-i( r_\mu \partial_\nu -r_\nu \partial_\mu) \,, \quad 
\partial_\mu \equiv \frac{\partial}{\partial r_\mu} \,,
\label{1}
\\ 
L_{\mu (n+1)}&=-L_{(n+1)\mu}=ir_{n+1}\partial_{\mu} \,, 
\label{2}
\end{align}
or more concisely 
\begin{align}
L_{ab}=-i( r_a \partial_b -r_b \partial_a )\,, \quad 
\partial_a\equiv \delta_{a\mu}\partial_{\mu} \,.
\label{3}
\end{align}
Noting that 
\begin{align}
\frac{\partial r_{n+1}}{\partial r_{\mu}}=-\frac{r_{\mu}}{r_{n+1}} \,, 
\label{4}
\end{align}
one can show that the operators in Eqs. (\ref{1}) and (\ref{2}) 
satisfy the O($n+1$) Lie algebra   
\begin{align}
[L_{ab}, L_{cd}]
=i ( \delta_{ac}L_{bd} -\delta_{bc}L_{ad} -\delta_{ad}L_{bc} +\delta_{bd}L_{ac}) \,.
\label{5}
\end{align}
In terms of the stereographic coordinates $(x_\mu)$ which are related to $(r_a)$ by 
\begin{align}
r_{\mu}=\frac{2 x_\mu}{1+x^2} \,, 
\quad 
r_{n+1}=\frac{1-x^2}{1+x^2} \,, \quad x \equiv \sqrt{x_\mu x_\mu} \,,
%x_{\mu}=\frac{r_\mu}{1+r_{n+1}} \,,
\label{6}
\end{align}
the angular momentum operators in Eqs. (\ref{1}) and (\ref{2}) are written as 
\cite{Ban} 
\begin{align}
L_{ab}=-i(r_a K_b^{\mu}-r_b K_a^{\mu}) \frac{\partial}{\partial x_\mu} \,,
\label{7}
\end{align}
where $K_a^{\mu}$ are the conformal Killing vectors 
\begin{align}
K_{\nu}^{\mu}=\frac{1+x^2}{2}\delta_{\mu\nu}-x_\mu x_\nu \,, 
\quad K_{n+1}^{\mu}=-x_\mu \,.
\label{8}
\end{align}
Equation (\ref{6}) illustrates a mapping from the flat space (or hyperplane) 
$\bar{\bf R}^{n}\equiv {\bf R}^{n} \cup \{\infty\}$ to $S_{1}^{n}$. 
The inverse mapping from $S_{1}^{n}$ to $\bar{\bf R}^{n}$ is then  
illustrated by $x_\mu=r_{\mu}/(1+r_{n+1})$. 
The Killing vectors $K_a^{\mu}$ satisfy the transversality condition 
\begin{align}
r_a K_a^{\mu}=0 \,.
\label{9}
\end{align}

Let $\hat{A}_{a}$ be a Yang--Mills field on $S^{n}_{1}$ which takes values in a 
Lie algebra ${\frak g}$; $\hat{A}_{a}$ can be expanded as  
$\hat{A}_{a}=\hat{A}_{a}^{i}T^{i}$ in 
terms of the Hermitian basis $\{ T^{i} \}$ of ${\frak g}$ which satisfy the algebra 
$  [T^{i}, T^{j}]=if^{ijk}T^{k} $ with the structure constants $f^{ijk}$. 
The normalization conditions $\mathrm{Tr}(T_i T_j)=\delta_{ij}$ are also put for convenience. 
One may regard $\hat{A}_{a}$ as a function of the independent variables $(r_{\mu})$. 
The Yang-Mills field $\hat{A}_{a}$ is assumed to live on the tangent space at a point 
$P(r_{\mu})$ on $S_1^n$ by imposing the transversality condition 
\begin{align}
r_a \hat{A}_{a}=0 \,.
\label{10}
\end{align}
This implies that one component of $(\hat{A}_{a})$, for instance $\hat{A}_{n+1}$, depends 
on the other components, such as $\hat{A}_{n+1}=-(r_\mu \hat{A}_{\mu})/r_{n+1}$. 
The differentiation of the condition (\ref{10}) with respect to $r_{\mu}$ is carried out 
to get 
\begin{align}
r_b \partial_{\mu} \hat{A}_b =-\hat{A}_{\mu}+
\frac{r_{\mu}}{r_{n+1}} \hat{A}_{n+1} \,.
\label{11}
\end{align}
Here Eq. (\ref{4}) has been used. Equation (\ref{11}) may be written as 
\begin{align}
r_b \partial_{a} \hat{A}_b =-\hat{A}_{a}+
\frac{r_{a}}{r_{n+1}} \hat{A}_{n+1} \,,
\label{12}
\end{align}
because, in the case $a=\mu$, it reduces to Eq. (\ref{11}) and, in the case 
$a=n+1$, both sides of Eq. (\ref{12}) vanish so that it gives rise to no additional 
conditions. 
(Note that $\partial_a$ is understood to be $\delta_{a\mu}\partial_{\mu}$.)

The infinitesimal gauge transformation of $\hat{A}_{a}$ is given by \cite{Ban}
\begin{align}
\delta_{\lambda} \hat{A}_a 
&= ir_b \mathcal{L}_{ba} \lambda =P_{ab}\hat{D}_{b} \lambda \,, 
\label{13}
\end{align}
where $\lambda$ is an infinitesimal gauge parameter taking values in ${\frak g}$, 
$\mathcal{L}_{ab}$ are covariantized angular momentum operators  
%(in the adjoint representation) 
%
\begin{align}
\mathcal{L}_{ab}&\equiv L_{ab}-[r_a \hat{A}_b -r_b \hat{A}_a, \;\;\, ]
\nonumber \\
&=-i(r_a \hat{D}_b -r_b \hat{D}_a) \,,
\label{14}
\end{align}
while $P_{ab}$ and $D_{a}$ are the tangential projection operator and the covariant 
derivative, respectively: 
\begin{align}
P_{ab} &\equiv \delta_{ab}-r_a r_b \,, 
\label{15}
\\ 
\hat{D}_a &\equiv \partial_a -i[ \hat{A}_a, \;\;\, ] \,.
\label{16}
\end{align}
The projection operator $P_{ab}$ in Eq. (\ref{13}) guarantees that the Yang--Mills field 
transformed according to the rule (\ref{13}), i.e.,  
$\hat{A}_a +\delta_{\lambda} \hat{A}_a$, lives on the same tangent space.

The field strength of $\hat{A}_{a}$ can be written in a manifestly 
O($n+1$)-covariant form \cite{JR,Ore,Ban}: 
\begin{align}
\hat{F}_{abc}&= i( L_{ab}\hat{A}_{c}-r_a [\hat{A}_{b}, \hat{A}_{c} ] \big)
\nonumber \\ 
&\,\quad +\mbox{cyclic permutations in $(a, b, c)$}
\nonumber \\ 
&= r_a \hat{F}_{bc} +r_b \hat{F}_{ca} +r_c \hat{F}_{ab} \,, 
\label{17}
\end{align}
where $\hat{F}_{ab}$ is defined by 
\begin{align}
\hat{F}_{ab}=\partial_a \hat{A}_b -\partial_b \hat{A}_a
-i [\hat{A}_{a}, \hat{A}_{b} ] \,. 
\label{18}
\end{align}
The gauge transformation of $\hat{F}_{ab}$ is found from Eq. (\ref{13}) 
to be 
\begin{align}
\delta_{\lambda} \hat{F}_{ab}&= -i[ \hat{F}_{ab}, \lambda] 
+r_a \bigg( \hat{D}_{b}+\frac{1}{r_{n+1}}\delta_{b (n+1)}
\bigg)(r_{\mu}\partial_{\mu} \lambda) 
\nonumber \\ 
&\quad \,-r_b \bigg( \hat{D}_{a}+\frac{1}{r_{n+1}}\delta_{a (n+1)}
\bigg)(r_{\mu}\partial_{\mu} \lambda) \,, 
\label{19} 
\end{align}
where Eqs. (\ref{4}), (\ref{10}) and (\ref{11}) have been used. 
This is an inhomogeneous transformation involving terms that are not manifestly 
O($n+1$) covariant. 
Since field strengths should transform homogeneously, 
one cannot take $\hat{F}_{ab}$ 
as the field strength of $\hat{A}_{a}$, even though $\hat{F}_{ab}$ looks like 
the field strength as far as one sees only Eq. (\ref{18}). 
Instead of $\hat{F}_{ab}$, the rank-3 tensor $\hat{F}_{abc}$ transforms 
homogeneously under the gauge transformation,  
\begin{align}
\delta_{\lambda} \hat{F}_{abc}= -i[ \hat{F}_{abc}, \lambda] \,.
\label{20}
\end{align}
Hence $\hat{F}_{abc}$ has the property of field strength. 
With the field strength $\hat{F}_{abc}$, the Yang--Mills action for $\hat{A}_{a}$ 
is given by 
\begin{align}
S_{\rm YM}=\int d\Omega\,\bigg[ -{1\over12} \mathrm{Tr}(\hat{F}_{abc} \hat{F}_{abc}) 
\bigg] \,. 
\label{21}
\end{align}
Here $d\Omega$ is an invariant measure on $S_1^n$ which is written in terms of 
the coordinates $(r_{\mu})$ as 
\begin{align}
d\Omega= \frac{1}{|r_{n+1}|} \prod_{\mu=1}^{n} dr_{\mu} \,. 
\label{22}
\end{align} 
Obviously, the action $S_{\rm YM}$ is gauge invariant.
The variation of $S_{\rm YM}$ can be calculated through integration by parts 
over $(r_{\mu})$, which is performed by taking into account the factor 
$|r_{n+1}|^{-1}$ contained in $d\Omega$.  
Using Eq. (\ref{4}) and noting that $S_1^{n}$ has no boundaries, we obtain 
\begin{align}
\delta S_{\rm YM}=\int d\Omega\,\mathrm{Tr}\bigg[  
\frac{i}{2}\delta\hat{A}_{c} \mathcal{L}_{ab}\hat{F}_{abc}
\bigg] \,.
\label{23}
\end{align}

\section{gauge-fixing conditions}

For studying quantum-theoretical structure of the Yang--Mills theory on 
the hypersphere $S_1^n$, 
it is necessary to consider the gauge-fixing procedure in the theory. 
We here focus our attention on discussing gauge-fixing conditions, 
before setting a suitable gauge-fixing term.

In a study of massless Euclidean QED on $S_1^4$, Adler proposed 
a gauge-fixing condition $iL_{ab} \hat{A}_b =\hat{A}_a$ \cite{Adl}. 
This is, of course, useful for making the analysis quite simple, 
and is also applicable to higher-dimensions as well as to the case of Yang--Mills theory. 
However, Adler's condition is unusual in the sense that 
it has an extra free index $a$ in comparison with the well-known Lorentz 
condition $\partial_{\mu} A_\mu=0\,$;  
for this reason, Adler's condition is not favorable to the ordinary 
first-order formalism of gauge-fixing \cite{NL,NO}. 
In this section, we shall prove that Adler's condition 
is equivalent to the condition $ir_{a}L_{ab} \hat{A}_b =0$ with no free indices. 
This condition, which has a form more similar to $\partial_{\mu} A_\mu=0$ 
than Adler's condition,
is essentially the same as 
the one used in Refs. \cite{Ore,BP} in a somewhat different context where  
the one-instanton background is present. 
Although Adler's condition (or its quadratic equivalent) 
was compared with another condition 
$ir_{a}L_{ab} \hat{A}_b =0$ in the literature \cite{Sho}, 
it seems that a complete proof of the equivalence has not been given yet.

To incorporate a gauge parameter $\alpha$ into the gauge-fixing conditions 
in a manifestly O($n+1$)-covariant manner, 
we introduce the Nakanishi--Lautrup (NL) field $\hat{B}$ on $S_1^{n}$ 
\cite{NL}. 
Then the above-mentioned conditions are generalized as in Eqs.~(\ref{24}) and 
(\ref{25}) given below. The generalization of Adler's condition, Eq.~(\ref{24}),  
is also unusual in the sense that it has a free index $a$. 
We now show the following.  
\\

\noindent
{\bf Proposition:}$\;\;$The following conditions (a) and (b) are equivalent. 
\begin{align}
\mbox{(a)}& \quad iL_{ab}\hat{A}_b +\alpha r_a \hat{B}=\hat{A}_a \,,
{}^{\footnotemark[1]}
\label{24} 
\\ 
\mbox{(b)}& \quad ir_a L_{ab}\hat{A}_b +\alpha \hat{B}=0 \,.
\label{25}
\end{align}
% 
%+++++++++++++++++++++++++++++++++++++++++++++++++++++++++++++++++++++++++++++
%
\footnotetext[1]{Using Eqs.~(10) and (12), we can readily show that 
$\hat{A}_a (iL_{ab}\hat{A}_b -\hat{A}_a)=0$, which is compatible with the condition 
(a). The equation derived here implies that 
the vector $(iL_{ab}\hat{A}_b -\hat{A}_a)$ is null or perpendicular to the vector 
$(A_{a})$ living in a tangent space of $S_1^{n}$. 
Whereas $(iL_{ab}\hat{A}_b -\hat{A}_a)$ can live in the same tangent space, 
the condition (a) requires that $(iL_{ab}\hat{A}_b -\hat{A}_a)$ is null or 
in the normal (or radial) direction, being perpendicular to $S_1^n$.}{\bf Proof:}$\;\;$ 
Consider the condition (a). Contracting Eq.~(\ref{24})  
by $r_a$ yields Eq.~ (\ref{25}), owing to the constraint  
$r_a r_a=1$ and the condition (\ref{10}). 
The condition (b) is thus derived from (a). 

Next we shall derive (a) from (b). 
Using Eq.~(\ref{12}), $r_a r_a=1$ and the condition (\ref{10}), 
one can rewrite Eq. (\ref{25}) as 
\begin{align}
\partial_b \hat{A}_{b} =\frac{1}{r_{n+1}}\hat{A}_{n+1}-\alpha \hat{B} \,.
\label{26}
\end{align}
This is an alternative form of the condition (b). 
Substituting Eqs.~(\ref{12}) and (\ref{26}) into 
\begin{align}
iL_{ab}\hat{A}_b = r_a \partial_b \hat{A}_{b}-r_b \partial_a \hat{A}_{b} 
\label{27}
\end{align}
leads us to Eq.~(\ref{24}).  
Consequently, the equivalence of the conditions (a) and (b) is established. 
%({\it Q.E.D.})
{\small $\:\Box$} 
\\

Since Eqs.~(\ref{24}) and (\ref{25}) hold simultaneously, 
we can eliminate the NL field $\hat{B}$ from Eq.~(\ref{24}) with the aid 
of Eq.~(\ref{25}), getting  
\begin{align}
iL_{ab}\hat{A}_b -\hat{A}_a =r_a (ir_c L_{cb}\hat{A}_b)  \,.
\label{28}
\end{align}
From this formula, we readily see that the condition $ir_a L_{ab}\hat{A}_b=0$ 
is equivalent to Adler's condition $iL_{ab}\hat{A}_b =\hat{A}_a$. 
Adding $\alpha r_a \hat{B}$ to the both sides of Eq.~(\ref{28}), we have  
\begin{align}
iL_{ab}\hat{A}_b -\hat{A}_a +\alpha r_a \hat{B} 
=r_a (ir_c L_{cb}\hat{A}_b +\alpha\hat{B})  \,. 
\label{29}
\end{align}
This formula converts the conditions (a) and (b) into each other. 
Equation~(\ref{29}) will also be useful for simplifying field equations.

\section{BRST symmetry and a gauge-fixing term}

It is known that gauge-fixing is neatly performed by considering the BRST invariance  
as a first principle \cite{KU,NO}. 
In the previous section, we have claimed that Adler's condition and 
its generalization, namely the condition (a), are not conventional owing to 
the presence of a free index. 
If one applies the BRST gauge-fixing procedure to the Yang--Mills theory on $S_1^n$, 
the condition (b) is much better to use than (a), though (a) and (b) are equivalent. 
The reason for choosing (b) is that it 
resembles the ordinary gauge-fixing condition $\partial_{\mu}A_{\mu}+\alpha B=0$ 
which has often been adopted in the BRST gauge-fixing procedure \cite{KU,NO}.  
With the condition (b), it is easy to apply the conventional 
procedure to the present case since there is no free index.

The BRST transformation $\boldsymbol{\delta}$ is defined for $\hat{A}_{a}$ by replacing 
$\lambda$ in Eq.~(\ref{13}) by the Faddeev--Popov (FP) ghost field $\hat{C}$ on $S_1^{n}$, 
\begin{align}
\boldsymbol\delta \hat{A}_a &= ir_b \mathcal{L}_{ba} \hat{C} =P_{ab}\hat{D}_{b} \hat{C} \,.
\label{30}
\end{align}
By putting the transformation rule 
\begin{align}
\boldsymbol\delta \hat{C}=\frac{i}{2}\{ \hat{C}, \hat{C} \} \,,
\label{31}
\end{align}
the nilpotency property $\boldsymbol{\delta}^{2}=0$ is guaranteed for  
$\hat{A}_{a}$ and $\hat{C}$. In particular, $\boldsymbol{\delta}^{2}\hat{A}_{a}=0$ 
is verified by using the property $P_{ac}P_{cb}=P_{ab}$.
In addition to $\hat{C}$, 
the FP anti-ghost field $\hat{\bar{C}}$ is introduced to satisfy 
\begin{align}
\boldsymbol\delta \hat{\bar{C}}=i\hat{B}\,, \quad 
\boldsymbol\delta \hat{B}=0 \,.
\label{32}
\end{align}
Consequently the nilpotency property $\boldsymbol{\delta}^{2}=0$ is still held  
after incorporating $\hat{\bar{C}}$ and $\hat{B}$. 
Needless to say, $\hat{C}$, $\hat{\bar{C}}$ and $\hat{B}$ are treated as 
functions of $(r_\mu)$.

With the BRST transformation and the relevant fields in hand,  
we can discuss the BRST gauge-fixing procedure for the Yang--Mills theory on $S_1^n$. 
To deal with the condition (b) in a BRST symmetric manner, we now take   
the sum of gauge-fixing and FP ghost terms written in the following 
form:$\,{}^{\footnotemark[2]}$ 
%
%
%%%%%%%%%%%%%%%%%%%%%%%%%%%%%%%%%%%%%%%%%%%%%%%%%%%%%%%%%%%%%%%%%%%%%%%%%%%%%%%%%%%%
%
\footnotetext[2]{It is possible to deal with Adler's gauge-fixing condition by taking   
$$ \tilde{S}_{\rm GF}= \int d\Omega \bigg\{ -i \boldsymbol\delta \mathrm{Tr}
\bigg[ \hat{\bar{C}}_a \bigg(i L_{ab} \hat{A}_b -\hat{A}_a 
+\frac{\alpha}{2} \hat{B}_a \bigg) \bigg] \bigg\} 
$$
instead of Eq.~(\ref{33}). 
Here $\hat{B}_a$ and $\hat{\bar{C}}_a$ are a NL field and a FP anti-ghost field of 
the vector type, respectively, satisfying 
$\boldsymbol\delta \hat{\bar{C}}_a=i\hat{B}_a,\;\boldsymbol\delta \hat{B}_a=0.$
The necessity of introducing the vector type of NL and FP anti-ghost fields  
is due to the fact that Adler's gauge-fixing condition has a free index. 
Although $\tilde{S}_{\rm GF}$ works well as a sum of gauge-fixing and FP ghost 
terms, it involves redundant 
degrees of freedom caused by the vectorial property of $\hat{B}_a$ and $\hat{\bar{C}}_a$. 
As a result, the discussion based on $\tilde{S}_{\rm GF}$ is fairly complicated. 
Also, in this case, the symmetry between a FP ghost field and a FP anti-ghost field is spoiled. 
This turns out to be a serious problem when one applies the superfield formalism 
\cite{BT}  
of the BRST and anti-BRST symmetries to  
the Yang--Mills theory on $S_1^n$ \cite{BD}. }
%
%%%%%%%%%%%%%%%%%%%%%%%%%%%%%%%%%%%%%%%%%%%%%%%%%%%%%%%%%%%%%%%%%%%%%%%%%%%%%%%%%%%%%
%
%
\begin{align}
S_{\rm GF}= \int d\Omega \bigg\{ -i \boldsymbol\delta \mathrm{Tr}
\bigg[ \hat{\bar{C}} \bigg(ir_a L_{ab} \hat{A}_b +\frac{\alpha}{2} \hat{B} \bigg) 
\bigg] \bigg\} \,. 
\label{33}
\end{align}
Since this is a BRST-coboundary term, the BRST invariance of $S_{\rm GF}$ is guaranteed 
due to the nilpotency of $\boldsymbol\delta$. 
In contrast, the BRST invariance 
of $S_{\rm YM}$ is clear from its gauge invariance. 
Carrying out the BRST transformation contained in the right-hand side of Eq.~(\ref{33}) 
and using the formula 
\begin{align}
r_a L_{ac}(r_b \mathcal{L}_{bc})=\frac{1}{2} L_{ab} \mathcal{L}_{ab} \,,
\label{34}
\end{align}
we have
\begin{align}
S_{\rm GF}= \int d\Omega\,  \mathrm{Tr}\bigg[ 
\hat{B} ir_a L_{ab} \hat{A}_b +\frac{\alpha}{2} \hat{B}^2 
-\frac{1}{2}i \hat{\bar{C}} L_{ab} \mathcal{L}_{ab} \hat{C}
\bigg] \,. 
\label{35}
\end{align}
Integrating by parts over $(r_{\mu})$ and using Eqs. (4), (10) and 
$r_a r_a=1$, we may rewrite Eq.~(\ref{35}) as 
\begin{align}
S_{\rm GF}& = \int d\Omega\,  \mathrm{Tr}\bigg[ 
-(ir_b L_{ba} \hat{B}) \hat{A}_a +\frac{\alpha}{2} \hat{B}^2 
+ (ir_b L_{ba} \hat{\bar{C}}) r_{c}  \mathcal{L}_{ca} \hat{C}
\bigg] 
\nonumber \\ 
& = \int d\Omega\,  \mathrm{Tr}\bigg[ 
-\big( ir_b L_{ba} \hat{B}-\{ ir_b L_{ba} \hat{\bar{C}}, \hat{C} \} \big) \hat{A}_a 
+\frac{\alpha}{2} \hat{B}^2 
+ (ir_b L_{ba} \hat{\bar{C}}) r_{c} L_{ca} \hat{C} \bigg] 
\label{36}
\end{align}
Further integration by parts for the ghost term in  Eq.~(\ref{36}) leads to 
\begin{align}
S_{\rm GF}= \int d\Omega\, \mathrm{Tr}\bigg[ 
-(ir_b L_{ba} \hat{B}) \hat{A}_a +\frac{\alpha}{2} \hat{B}^2 
-{1\over2}i (\mathcal{L}_{ab} L_{ab} \hat{\bar{C}}) \hat{C}
\bigg] \,, 
\label{37}
\end{align}
with use of the formula
\begin{align}
r_a \mathcal{L}_{ac}(r_b L_{bc})=\frac{1}{2} \mathcal{L}_{ab} L_{ab} \,. 
\label{38}
\end{align}
Quantization of the fields $\hat{A}_a$, $\hat{C}$, $\hat{\bar{C}}$ and $\hat{B}$ 
is performed based on the total action 
\begin{align}
S=S_{\rm YM}+S_{\rm GF} 
\label{39}
\end{align}
in a systematic way; thereby one may see quantum-theoretical structure of the 
Yang--Mills theory on $S_1^n$. Details of the quantization will be discussed elsewhere.

\section{Field equations}

From the total action $S$, we can derive the Euler--Lagrange equation for each field, 
\begin{align}
&{i\over2} \mathcal{L}_{ab} \hat{F}_{abc}
=ir_b L_{bc} \hat{B}-\{ ir_b L_{bc} \hat{\bar{C}}, \hat{C} \} \,,
\label{40} 
\\ 
& ir_a L_{ab} \hat{A}_b +\alpha \hat{B}=0 \,, 
\label{41} 
\\
& L_{ab} \mathcal{L}_{ab} \hat{C}=0 \,, 
\label{42} 
\\
& \mathcal{L}_{ab} L_{ab} \hat{\bar{C}}=0 \,. 
\label{43}
\end{align}
The field equation (\ref{40}), which is directly found from Eqs.~(\ref{23}) 
and (\ref{36}), is just the Yang--Mills equation on $S_1^n$ including 
the additional terms due to $S_{\rm GF}$. 
By contracting Eq.~(\ref{40}) by $r_c$, both sides of this equation vanish,    
implying that Eq.~(\ref{40}) satisfies the transversality condition. 
If the right-hand side of Eq.~(\ref{40}) is put to be zero, it reduces to 
the Yang--Mills equation obtained by Jackiw and Rebbi \cite{JR}. 
Equation (\ref{41}), which is easily read-off from Eq.~(\ref{35}), is precisely 
the condition (b). 
The field equations (\ref{42}) and (\ref{43}) follow  from  
Eqs. (\ref{35}) and (\ref{37}), respectively. 
From Eq.~(\ref{40}), the field equation governing the motion of $\hat{B}$ 
is obtained as 

\begin{align}
\mathcal{L}_{ab} L_{ab} \hat{B}=\{ L_{ab} \hat{\bar{C}}, \mathcal{L}_{ab} \hat{C} \}
\label{44}
\end{align}
after using Eq.~(\ref{43}) and $r_a r_a=1$. 
In the Abelian case, using the commutation relations (\ref{5}) and 
$[L_{ab} ,r_c]=-i(r_a \delta_{bc}-r_b \delta_{ac})$, as well as Eqs.~(\ref{29}) and 
(\ref{41}), we can rewrite Eq.~(\ref{40}) as 
\begin{align}
L_{ab}L_{ab}\hat{A}_c +2(n-2)\hat{A}_c +4\alpha r_c \hat{B}
+2i(1-\alpha) r_b L_{bc} \hat{B}=0 \,.
\label{45}
\end{align}
Even in the Feynman gauge $\alpha=1$, the NL field $\hat{B}$ still remains in 
the third term of Eq.~(\ref{45}). This is a difference between the present case 
and the ordinary case formulated on the flat space. Contracting Eq.~(\ref{45}) by $r_c$ 
reproduces Eq.~(\ref{41}) by virtue of the presence of the third term, 
which fact demonstrates compatibility of the field equation (\ref{45}) and 
the gauge condition (\ref{41}). In the particular situation of $\hat{B}=0$, 
Eq.~(\ref{45}) in four dimensions reduces to the field equation found by Adler \cite{Adl}.

The field equations Eqs.~(\ref{40})--(\ref{44}) 
have manifestly O($n+1$) covariant or invariant forms.
They are the  spherical 
analogues of the field equations on the flat space given in the literature 
\cite{KO,NO}:   
\begin{align}
& D_{\mu} F_{\mu\nu}=\partial_{\nu} B-\{ \partial_{\nu} \bar{C}, C \} \,, 
\label{46} 
\\
& \partial_{\mu}A_{\mu}+\alpha B=0 \,, 
\label{47} 
\\ 
& \partial_{\mu} D_{\mu} C=0 \,, 
\label{48}
\\
& D_{\mu} \partial_{\mu} \bar{C}=0 \,, 
\label{49}
\\
& D_{\mu}\partial_{\mu} B=\{ \partial_{\mu} \bar{C}, D_{\mu} C \} \,, 
\label{50}
\end{align}
where 
\begin{align}
D_\mu & \equiv\partial_\mu -i[ A_\mu, \;\;\, ] \,, 
\label{51} 
\\ 
F_{\mu\nu} & \equiv\partial_\mu A_\nu -\partial_\nu A_\mu -i[A_\mu, A_\nu] \,. 
\label{52}
\end{align}
Comparing Eqs.~(\ref{40})--(\ref{44}) with Eqs.~(\ref{46})--(\ref{50}), 
we see that $L_{ab}$ and $\mathcal{L}_{ab}$ correspond to $\partial_{\mu}$ 
and $D_{\mu}$, respectively. 
In the Yang--Mills theory on $S_1^n$, $L_{ab}$ and $\mathcal{L}_{ab}$ 
are more fundamental than $\partial_{\mu}$ and $D_{\mu}$. 
This can be understood from the fact that 
translations on a plane are realized as rotations on a sphere, 
so that usual derivatives are replaced by angular derivations 
when one discusses on a sphere. 
It is possible to establish the correspondence between 
Eqs.~(\ref{40})--(\ref{44}) 
and Eqs.~(\ref{46})--(\ref{50}) through the following discussion.

The Yang--Mills field $\hat{A}_{a}$ on the hypersphere $S_1^{n}$ and 
the conventional one $A_\mu$ on the flat space $\bar{\bf R}^{n}$ is related  
by the conformal Killing vectors $K_a^{\mu}$ 
\cite{Ban}:  

\begin{align}
\hat{A}_a(r) =K_a^{\mu} A_{\mu}(x) \,.
\label{53}
\end{align}
The field strength $\hat{F}_{abc}$ can be expressed as \cite{Ban} 
\begin{align}
\hat{F}_{abc}=\big(r_a K_b^{\mu} K_c^{\nu} +r_b K_c^{\mu} K_a^{\nu} 
+r_c K_a^{\mu} K_b^{\nu} \big) F_{\mu\nu} \,, 
\label{54}
\end{align}
with $F_{\mu\nu}$ defined by Eq.~(\ref{52}). 
Similarly to Eq.~(\ref{53}), the fields $\hat{B}$, $\hat{C}$ and $\hat{\bar{C}}$  
on $S_1^{n}$ are related to the corresponding fields $B$, $C$ and $\bar{C}$ on 
$\bar{\bf R}^{n}$ by 
\begin{align}
\hat{B}(r)=\bigg(\frac{1+x^2}{2} \bigg)^{\!2} B(x) \,, \quad
\hat{C}(r)=\frac{1+x^2}{2} C(x) \,, \quad 
\hat{\bar{C}}(r)=\frac{1+x^2}{2} \bar{C}(x) \,. 
\label{55}
\end{align}
These relations may be understood from a supersphere formulation of 
the Yang--Mills theory on $S_1^n$ \cite{BD}.  
From Eq.~(\ref{7}), it follows that 
\begin{align}
ir_b L_{ba}=K_{a}^{\mu}\partial_\mu \,, \quad 
\partial_\mu \equiv \frac{\partial}{\partial x_\mu} \,, 
\label{56}
\end{align}
where $r_a r_a=1$ and the condition (\ref{9}) has been used. 
A covariantized version of Eq.~(\ref{56}) is also satisfied owing to Eq.~(\ref{10}), 
\begin{align}
ir_b \mathcal{L}_{ba}=K_{a}^{\mu} D_{\mu} \,,
\label{57}
\end{align}
with $D_{\mu}$ defined by Eq.~(\ref{51}). 
We now rewrite the field equations (\ref{40})--(\ref{44}) in terms of 
the relevant fields on $\bar{\bf R}^{n}$. 
To this end, the following formulas are particularly useful: 
\begin{align}
&K_a^{\mu} K_a^{\nu}= \bigg(\frac{1+x^2}{2} \bigg)^{\!2} \delta_{\mu\nu} \,, 
\label{58} 
\\ 
&K_a^{\mu} \partial_{\mu} K_a^{\nu}=(2-n) \bigg(\frac{1+x^2}{2} \bigg) x_{\nu} \,, 
\label{59}
\\ 
&K_a^{\rho}\partial_{\rho}(K_a^\mu K_b^\nu -K_a^\nu K_b^\mu)
\nonumber \\ 
&=(4-n) \frac{1+x^2}{2} (x_\mu K_b^\nu -x_\nu K_b^\mu ) \,.
\label{60}
\end{align}
Substituting Eqs.~(\ref{53})--(\ref{55}) into Eqs.~(\ref{40})--(\ref{44}) and 
using the formulas (\ref{58})--(\ref{60}), 
together with $r_a r_a=1$ and Eqs.~(\ref{9}), (\ref{10}), 
(\ref{34}), (\ref{38}), (\ref{56}) and (\ref{57}), at appropriate stages, 
we can write Eqs.~(\ref{40})--(\ref{44}) in terms of the stereographic coordinates: 
\begin{align}
& D_{\mu}F_{\mu\nu}
+\frac{2(4-n)}{1+x^2}\, x_{\mu} F_{\mu\nu}
\nonumber 
\\ 
&=\partial_{\nu} B +\frac{4x_{\nu}}{1+x^2} B 
-\bigg\{ \partial_{\nu} \bar{C} +\frac{2x_{\nu}}{1+x^2} \bar{C}, C \bigg\} \,, 
\label{61} 
\\ 
& \partial_{\mu} A_{\mu} +\frac{2(2-n)}{1+x^2}\, x_{\mu} A_{\mu} +\alpha B=0 \,, 
\label{62}
\\
& \partial_{\mu} D_{\mu}C +\frac{2x_{\mu}}{1+x^2} 
\{(3-n)D_{\mu}C +\partial_{\mu}C \}
+\frac{2(4-n)x^2 +2n}{(1+x^2)^2} C=0 \,, 
\label{63}
\\
& D_{\mu} \partial_{\mu}\bar{C} +\frac{2x_{\mu}}{1+x^2} 
\{(3-n)\partial_{\mu}\bar{C} +D_{\mu}\bar{C} \}
+\frac{2(4-n)x^2 +2n}{(1+x^2)^2} \bar{C}=0 \,, 
\label{64}
\\
& D_{\mu} \partial_{\mu} B +\frac{2x_{\mu}}{1+x^2} 
\{(4-n)\partial_{\mu}B +2D_{\mu}B \}
+\frac{4(6-n)x^2 +4n}{(1+x^2)^2} B 
\nonumber \\ 
& =\bigg\{ \partial_{\mu} \bar{C} +\frac{2x_{\mu}}{1+x^2} \bar{C}, \,
D_{\mu} C +\frac{2x_{\mu}}{1+x^2} C \bigg\} \,. 
\label{65}
\end{align} 
Equation (\ref{62}) may be regarded as a combination of the ordinary gauge-fixing  
condition $\partial_\mu A_\mu+\alpha B=0$ and the Fock--Schwinger gauge condition 
$x_\mu A_\mu=0$ \cite{Sch}, involving an extra factor $(1+x^2)^{-1}$. 
In the case of $\alpha=0$, 
Eq. (\ref{62}) is essentially a higher-dimensional analogue of the gauge-fixing condition 
proposed by Ore \cite{Ore}. 
An analogue of Eq. (\ref{63}) is also seen in Ref. \cite{Ore}. 
In the present paper, however, Eqs. (\ref{62}) and (\ref{63}) have been derived 
systematically by considering the BRST symmetry and using the conformal Killing vectors.

We now make a replacement of $r_a$ by $r_a /R$ to explicitly incorporate the radius 
$R$ of a hypersphere into the field equations. Thereby, the hypersphere $S_1^n$ 
is scaled to be $S_{R}^n$ characterized by $r_a r_a =R^2$, and the Yang--Mills theory in 
question is formulated on $S_{R}^n$. 
By taking the limit $R\rightarrow \infty$, the hypersphere $S_R^n$ approximates 
to the flat space, i.e. $S_{\infty}^n=\bar{\bf R}^{n}$. 
To maintain the relation between two coordinates systems $(r_\mu)$ and $(x_{\mu})$  
even in the large radius limit, $x_\mu$ has to be replaced by $x_\mu /R$. Correspondingly,
the derivative gets replaced as $\partial_\mu\mapsto R\partial_\mu$. 
Accordingly, the relevant fields on the flat space have to be modified appropriately.
For instance, to maintain homogeneity of the changes, $A_\mu \mapsto RA_\mu$, leading
to $F_{\mu\nu}\mapsto R^2 F_{\mu\nu}$. Any other transformation for $A_\mu$ would lead
to different transformations for the derivative piece and the commutator piece in $F_{\mu\nu}$
leading to a lack of homogeneity. Similarly, the other fields are modified as,  
$B \mapsto R^2 B$, $C \mapsto R C$ and 
$\bar{C} \mapsto R \bar{C}$. After making the replacement in 
Eqs.~(\ref{61})--(\ref{65}), one can take the limit $R\rightarrow \infty$ to  
see that Eqs.~(\ref{61})--(\ref{65}) reduce to Eqs.~(\ref{46})--(\ref{50}), 
respectively; the irrelevant terms in Eqs.~(\ref{61})--(\ref{65}) 
vanish on taking the limit. 
Therefore Eqs.~(\ref{46})--(\ref{50}) are recognized to be the large radius limit 
of Eqs.~(\ref{61})--(\ref{65}).

We can also express Eq.~(\ref{45}) as 
an Abelian field equation for $A_{\mu}$ and $B$ written 
in terms of the stereographic coordinates. 
After making the replacement mentioned above in this equation, 
the term corresponding to the third term of Eq.~(\ref{45}) vanishes 
in the limit $R\rightarrow \infty$. Hence, in the Feynman gauge $\alpha=1$, 
the NL field $B$ is completely removed and 
the Abelian field equation turns out to be 
the Laplace equation $\partial_{\mu}\partial_{\mu} A_{\nu}=0$.

\section{Conclusions}
After briefly reviewing a manifestly O($n+1$)-covariant formulation of 
the Yang--Mills theory on the hypersphere $S_1^n$, 
we have discussed a gauge-fixing procedure for this theory by considering 
the BRST invariance as a first principle. 
It was stressed that although Adler's gauge-fixing condition is useful for concrete 
analyses, it is unfavorable for the BRST gauge-fixing procedure 
owing to the presence of an extra free index. 
Instead of Adler's condition, we proposed a suitable gauge-fixing condition which is 
equivalent to a generalization of Adler's condition, but has no extra free indices. 
A complete proof of the equivalence was given by introducing the NL field.

Having obtained the suitable gauge-fixing condition, the BRST gauge-fixing 
procedure was applied to the Yang--Mills theory on  $S_1^n$,  
toward investigating its quantum-theoretical properties. 
The gauge-fixing and FP ghost terms, as well as the Yang--Mills action, were written 
in manifestly O($n+1$)-invariant forms with the aid of the angular momentum operators 
and their gauge-covariantized versions. 
Consequently, the field equation for each of the Yang--Mills, NL,   
FP ghost, and FP anti-ghost fields was derived in a manifestly O($n+1$) 
covariant or invariant form. 
All the field equations were also written in terms of the stereographic coordinates 
using the conformal Killing vectors which act like a metric in translating formulas
from the local coordinates $(r_\mu)$ to the stereographic coordinates $(x_\mu)$  
and vice-versa. 
Then it was shown that these equations   
reduce to corresponding field equations defined on the $n$-dimensional 
flat space, in the limit where the radius of the hypersphere is taken very large.

In quantizing the relevant fields in the Yang--Mills theory on the hypersphere $S_R^n$, 
one can expect that infrared divergences are automatically regularized,  
because $S_R^n$ is a bounded space with a maximum length.  
For this reason, the Yang--Mills theory on $S_R^n$ would be appropriate  
for studying properties of quantum chromodynamics (QCD) at a low-energy 
region, such as the gluon condensate of dimension 2 \cite{dim2}. 
Also, the BRST symmetry discussed in this paper may be suitable for 
the geometric approach to non-Abelian chiral anomalies with the use of 
decent equations and BRST cohomology \cite{SZ},   
because, there, space-time is assumed to be essentially the hypersphere $S_R^n$.  
In such an approach, non-Abelian chiral anomalies will be found in association 
with the axial anomaly (or singlet anomaly) evaluated on a hypersphere \cite{NS,Ban}.

\begin{acknowledgements}
One of the authors (S.D.) would like to thank Prof. K. Fujikawa for his encouragements.
Also, R.B. thanks the Nihon University for providing support and the members of the 
Institute of Quantum Science for their gracious hospitality during his visit 
when this work was started. 
% This work was supported in part by the Nihon University Research Grant.
\end{acknowledgements}

\end{document}